\documentstyle[aps,amsmath,axodraw]{revtex}  

\begin{document}        

\baselineskip 14pt
\title{An Improved Determination of the Fermi Coupling Constant, $G_F$}

\author{Robin G. Stuart}

\address{Randall Laboratory of Physics, University of Michigan\\
                  Ann Arbor, Michigan 48109-1120}

\maketitle              

\begin{abstract}        
Over 40 years after the calculation of the 1-loop QED corrections to
the muon lifetime, new theoretical developments have made it possible
to obtain an analytic expression for the complete 2-loop QED
contributions in the Fermi theory. The exact result for the effects
of virtual and real photons, virtual electrons, muons and hadrons as
well as $e^+e^-$ pair creation is
\[
\Delta\Gamma_{{\rm QED}}^{(2)}=
      \Gamma_0\left(\frac{\alpha}{\pi}\right)^2
        \left(\frac{156815}{5184}
             -\frac{1036}{27}\zeta(2)
                         -\frac{895}{36}\zeta(3)
                         +\frac{67}{8}\zeta(4)
                              +53\zeta(2)\ln2-(0.042\pm0.002)\right)
 \]
where $\Gamma_0$ is the tree-level width. This eliminates the theoretical
error in the extracted value of the Fermi coupling constant, $G_F$, which
was previously the source of the dominant uncertainty. The new value is
\[
G_F=(1.16637\pm0.00001)\times 10^{-5}\,{\rm GeV}^{-2}.
\]
The overall error has been roughly halved and is now entirely
experimental. Several experiments are planned for the next generation of
muon lifetime measurements and these can proceed unhindered by
theoretical uncertainties.
\end{abstract}   	

\section{Introduction}

The three fundamental input parameters that enter into all calculations
of electroweak physics are the electromagnetic coupling constant,
$\alpha$, the Fermi coupling constant, $G_F$, and the mass of the $Z^0$
boson, $M_Z$. Their current best values, along with their absolute
and relative errors are \cite{PDG,LEPEWWG}
\begin{xxalignat}{2}
{}\qquad\qquad\qquad\qquad
\alpha&=1/(137.0359895\pm0.0000061)&  (0.045\,{\rm ppm})& \\
G_F&=(1.16639\pm0.00002)\times10^{-5}\,{\rm GeV^{-2}}& (17\,{\rm ppm})& \\
M_Z&=91.1867\pm0.0021\,{\rm GeV}&           (23\,{\rm ppm})&
\end{xxalignat}

In the mid-80's,
just before the turn on of LEP, a CERN report concluded that the error
on $M_Z$ would be $\pm50$\,MeV or 550\,ppm and that ``A factor of 2--3
improvement can be reached with a determined effort'' \cite{CERN}.
It was thus generally believed that the error on $M_Z$
represented the limiting factor in the precision with which theoretical
predictions could be made. The situation has changed adiabatically,
however, and the relative error on $M_Z$ now approaches that of $G_F$.

Since LEP, as a machine, is not a radically new design, the lesson
that we should take is that it is extremely difficult to
predict the accuracy with which physical quantities will be measured,
even in the
relatively short term, and that one should constantly strive to reduce
such errors to the minimum level consistent with the available
technology. The possibility of precision physics at a muon
collider serves to emphasize this point.

With this in mind, and given the great cost and effort that was
expended in reducing the error on $M_Z$ to its current value,
it is reasonable to look again at $G_F$ and see what
is required to reduce its error to a level where it can never become an
obstacle limiting the accuracy with which theoretical predictions can be
made.

$G_F$ is extracted from the measured value of the muon lifetime,
$\tau_\mu=(2.19703\pm0.00004)\,\mu$s \cite{PDG}
and on the experimental side this
is currently the source of the dominant error.
New experiments are planned at the
Brookhaven National Laboratory, the Paul Scherrer Institute and
the Rutherford-Appleton Laboratory and
it is likely that the uncertainty on $G_F$ from this
source will be reduced to somewhere in the range 0.5--1\,ppm.

Most of the work reported here appears in
ref.s\cite{muonhad,muonprl}.

\section{The Fermi Coupling Constant}

As given above the current relative error on the Fermi constant is
$\delta G_F/G_F=1.7\times10^{-5}$. Of this $0.9\times10^{-5}$ is
experimental and $1.5\times10^{-5}$ is theoretical being an estimate
of unknown 2-loop QED corrections.

$G_F$ is related to the measured muon lifetime, $\tau$, by the formula
\begin{equation}
\frac{1}{\tau_\mu}\equiv\Gamma_\mu=\Gamma_0(1+\Delta q).
\label{eq:QEDcorr}
\end{equation}
where
\begin{equation}
\Gamma_0=\frac{G_F^2 m_\mu^5}{192\pi^3}
\label{eq:Gamma0}
\end{equation}
as calculated using the Fermi theory in which the weak interactions
are described by a contact interaction. $\Delta q$ encapsulates the
higher order QED corrections and may written as a perturbation series
in $\alpha_r=e_r^2/(4\pi)$, the renormalized electromagnetic
coupling constant. Thus
\begin{equation}
\Delta q=\sum_{i=0}^\infty\Delta q^{(i)}
\label{eq:DeltaqSeries}
\end{equation}
in which the index $i$ gives the power of, $\alpha_r$ that appears in
$\Delta q^{(i)}$. Note that Eq.(\ref{eq:QEDcorr}) differs from the
usual formula\cite{PDG} in ways that begin to become important at the
part-per-million level.
It is known\cite{KinoSirl,Berman} that
\begin{eqnarray}
\Delta q^{(0)}&=&-8x-12x^2\ln x+8x^3-x^4\\
\Delta q^{(1)}&=&\left(\frac{\alpha_r}{\pi}\right)
                 \left(\frac{25}{8}-3\zeta(2)\right)
               +{\cal O}(\alpha_r x\ln x)
\end{eqnarray}
where $x=m_e^2/m_\mu^2$ and
$\zeta$ is the Riemann zeta function with $\zeta(2)=\pi^2/6$.
That the $\Delta q^{(i)}$ remain finite in the limit $m_e\rightarrow0$
is a consequence of the Kinoshita-Lee-Nauenberg
theorem\cite{KinoshitaLeeNaue} whose discovery was largely prompted by
this particular observation.

Although the Fermi theory is not renormalizable, the $\Delta q^{(i)}$ can
be shown\cite{BermSirl} to be finite for all $i$.
This remarkable feature follows from the fact that the $V-A$ interaction
is invariant under a Fierz
rearrangement that interchanges the wavefunctions of the electron and
the muon neutrino.
Thus Fermi theory is equivalent to an effective
theory in which the muon and electron occupy the same fermion current
in the weak interaction lagrangian.
After fermion mass renormalization is performed the divergences in the
vector part of this current are independent of the fermion mass and hence
cancel in exactly the way they would for QED. The lagrangian of Fermi
theory is invariant under the transformations
$\psi_e\rightarrow\gamma_5\psi_e$ and
$m_e\rightarrow -m_e$. The QED corrections to the axial vector of the
part can thus be obtained from the those of the
vector part by changing the sign of the electron mass and hence
are finite as well. Moreover the two sets of corrections
are equal in the limit
$m_e\rightarrow0$ and, in that case, calculations need only be performed
using the vector part of the Fermi interaction. This conclusion
holds under any regularization prescription and avoids the complications
associated with the use of $\gamma_5$ in dimensional
regularization\cite{dimreg}.

The foregoing discussion does not apply to the $\beta$-decay of the
neutron where
the Fierz rearrangement generates scalar and pseudoscalar terms
that bear no resemblance to QED and the radiative corrections are
consequently not finite.

\section{The 2-loop QED Corrections to the Muon Lifetime}

The complete 2-loop QED corrections to the muon lifetime require
the calculation of matrix element for the processes,
$\mu^-\rightarrow\nu_\mu e^-\bar\nu_e$,
$\mu^-\rightarrow\nu_\mu e^-\bar\nu_e\gamma$,
$\mu^-\rightarrow\nu_\mu e^-\bar\nu_e\gamma\gamma$ and
$\mu^-\rightarrow\nu_\mu e^-\bar\nu_e e^+e^-$ with up to two
virtual photons. All processes contain infrared (IR) divergences coming
from either virtual
photons, soft bremsstrahlung or both. The cancellation of IR
divergences occurs between the various processes but this complication
may be avoided by exploiting cutting relations and calculating the 2-loop
corrections as imaginary parts of 4-loop diagrams, some of which are
shown in Fig.s 2 and 3. In these Feynman diagrams thick
lines represent a muon and the thin lines represent either the
electron or the neutrinos all of which are taken to be massless. Since
the external muon is on-shell any cut passing through a muon line will
vanish and the only cuts contributing to the imaginary part
are precisely the ones that generate the diagrams appearing in
the calculation of muon decay.

Recursion relations \cite{parialint} obtained by integration-by-parts
were first applied to reduce all dimensionally regularized integrals
to a small set of relatively simple integrals.
The well-behaved primitive integrals were then calculated by
taking the external muon momentum, $q$, off mass shell
to obtain expressions as power series in $x=-q^2/m_\mu^2$ and
logarithms of $x$ using well-established large mass expansion
techniques\cite{largemass}.
As the large mass expansion proceeds many terms, such as those
that are topologically tadpoles, can be immediately discarded
since they do not give rise to imaginary parts.
Since the final result is required for $x=1$ the complete
series must be summed which can now be done
in closed form in terms of polygamma functions and certain classes
of multiple nested sums \cite{sfunctions}.

All diagrams were calculated in a general covariant gauge
for the photon field and exact cancellation in the final result of the
dependence on the gauge parameter was demonstrated.

\subsection{Hadronic Contributions}

\begin{figure}
\begin{center}
\begin{picture}(62,120)(0,0)
\ArrowLine(0,0)(14,40)   \Vertex(14,40){1}
\ArrowLine(14,40)(21,60)
\ArrowLine(21,60)(14,80) \Vertex(14,80){1}
\ArrowLine(14,80)(0,120)
\ArrowLine(42,120)(23,62)
\ArrowLine(23,62)(62,120)
\PhotonArc(21,60)(21.08,109.29,250.71){3}{6.5}
\GOval(-1.08,60)(5,5)(0){0.4}
\Text(8,0)[bl]{$\mu^-$}
\Text(5,122)[tl]{$e^-$}
\Text(38.5,120)[tr]{$\bar\nu_e$}
\Text(71,116)[tr]{$\nu_\mu$}
\Text(21,-3)[tl]{(a)}
\end{picture}
\qquad
\begin{picture}(62,120)(0,0)
\ArrowLine(0,0)(5.25,15)       \Vertex(5.25,15){1}
\ArrowLine(5.25,15)(15.75,45)  \Vertex(15.75,45){1}
\ArrowLine(15.75,45)(21,60)
\ArrowLine(21,60)(0,120)
\ArrowLine(42,120)(23,62)
\ArrowLine(23,62)(62,120)
\PhotonArc(10.5,30)(15.89,70.71,250.71){3}{6.5}
\GOval(-4.65,34.80)(5,5)(0){0.4}
\Text(-7,65)[bl]{1} \Text(-7,60)[bl]{--} \Text(-7,53)[bl]{2}
\Text(21,-3)[tl]{(b)}
\end{picture}
\qquad
\begin{picture}(62,120)(0,0)
\ArrowLine(0,0)(21,60)
\ArrowLine(21,60)(15.75,75)     \Vertex(15.75,75){1}
\ArrowLine(15.75,75)(5.25,105)  \Vertex(5.25,105){1}
\ArrowLine(5.25,105)(0,120)
\ArrowLine(42,120)(23,62)
\ArrowLine(23,62)(62,120)
\PhotonArc(10.5,90)(15.89,109.29,289.29){3}{6.5}
\GOval(-4.65,85.2)(5,5)(0){0.4}
\Text(-7,65)[bl]{1} \Text(-7,60)[bl]{--} \Text(-7,53)[bl]{2}
\Text(21,-3)[tl]{(c)}
\end{picture}
\qquad
\begin{picture}(62,120)(0,0)
\ArrowLine(0,0)(10.5,30)    \Vertex(10.5,30){1}
\ArrowLine(10.5,30)(21,60)
\ArrowLine(21,60)(0,120)
\ArrowLine(42,120)(23,62)
\ArrowLine(23,62)(62,120)
\Line(6.0,27.83)(15.01,32.17)
\Line(8.33,34.51)(12.67,25.50)
\Text(0,65)[bl]{1} \Text(0,60)[bl]{--} \Text(0,53)[bl]{2}
\Text(16,27)[l]{$\delta m_\mu$}
\Text(21,-3)[tl]{(d)}
\end{picture}
\\
\end{center}
\caption{Hadronic contributions to muon decay after Fierz
           rearrangement of the contact interaction.}
\end{figure}
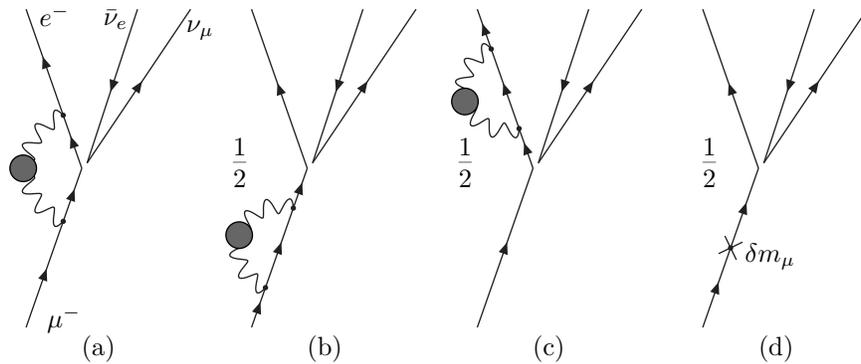

Hadronic effects enter $\tau_\mu$ at the 2-loop level through the
diagrams shown in Fig.1. The shaded blob represents the hadronic vacuum
polarization of the photon. The hadronic contribution can be calculated
in the usual way using dispersion relations but, in contrast to other
well-known situations for which such effects have been calculated
\cite{GourdeRa,KnieKrawKuhnStua}, the momenta of
the external fermions, to which the virtual photon is attached, is not
fixed. Here the electron participates in the phase-space integration
which complicates matters somewhat.

Hadronic contributions are always afflicted to some degree by an
uncertainty that arises from the experimental error on the measured
cross-section
$\sigma_{{\rm had}}\equiv\sigma(e^+e^-\rightarrow{\rm hadrons})$.
If this uncertainty turned out to be large there would be little
point proceeding with the perturbative calculation of the 2-loop
QED contributions to the muon lifetime.

The shift induced in the inverse lifetime, $\Gamma_\mu$, of the muon
is given as a convolution integral \cite{muonhad}
\begin{equation}
\Delta\Gamma_{\rm had}=
\frac{\alpha_r}{3\pi}\int_{4\rho}^\infty\frac{dz}{z}R(m_\mu^2 z)
                       \,\Delta\Gamma(z)
\label{eq:hadcorrz}
\end{equation}
over the hadronic spectrum,
$R(q^2)\equiv\sigma_{\rm had}/\sigma_{\rm point}$,
and in which $\rho=m_\pi^2/m_\mu^2=1.61395...$
The convolution kernel, $\Delta\Gamma(z)$ is obtained exactly as an
analytic function. When the integral
is performed using actual hadronic data the result is
\begin{equation}
\Delta\Gamma_{{\rm had}}=-\Gamma_0\left(\frac{\alpha_r}{\pi}\right)^2
                            (0.042\pm0.002)
\end{equation}
which includes a rather conservative estimate of the hadronic
uncertainty. Still the latter amounts to only 2 parts in $10^8$ and
so is well under control.

The integral (\ref{eq:hadcorrz}) can be used to obtain an expression
for the contribution from diagrams where the hadronic vacuum
polarization has been replaced by muon loop by setting
\begin{equation}
R(m_\mu^2z)=\left(1+\frac{2}{z}\right)\sqrt{1-\frac{4}{z}}.
\end{equation}
which gives
\begin{eqnarray}
\Delta\Gamma_{{\rm muon}}&=&\Gamma_0\left(\frac{\alpha_r}{\pi}\right)^2
\left(\frac{16987}{576}-\frac{85}{36}\zeta(2)
                        -\frac{64}{3}\zeta(3)\right)\\
                         &=&-\Gamma_0\left(\frac{\alpha_r}{\pi}\right)^2
                            0.0364333.
\end{eqnarray}
The result agrees with that subsequently obtained by perturbative
methods. The effect of tau loops can be obtained in a similar way and,
as expected on the
basis of the decoupling theorem, is very small.

\subsection{Photonic Corrections}

Examples of photonic diagrams which when cut give rise to contributions
to the muon lifetime at ${\cal O}(\alpha^2)$ are shown in Fig.2.

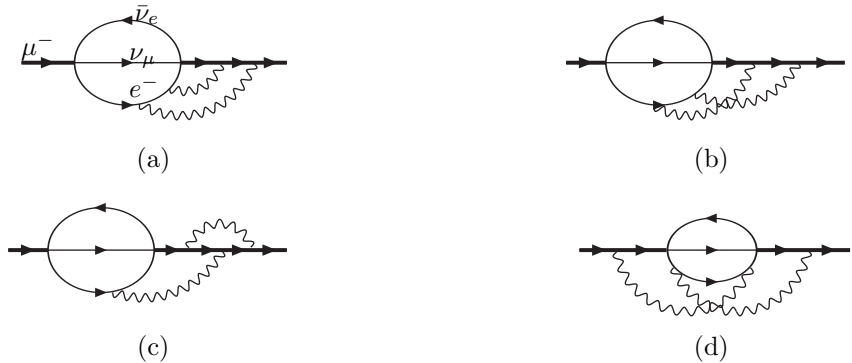
\begin{figure}
\begin{center}
\hfill
\begin{picture}(120,70)(0,0)
 \SetWidth{1.5}
 \Line(20,20)(40,20)
 \Text(20,23)[bl]{$\mu^-$}
 \Line(80,20)(120,20)
 \SetWidth{0.8}
 \ArrowLine(20,20)(40,20)
 \ArrowLine(80,20)(96,20)
 \ArrowLine(96,20)(108,20)
 \ArrowLine(108,20)(120,20)
 \SetWidth{0.5}
 \ArrowLine(40,20)(80,20)
 \Text(60.8,21)[bl]{$\nu_\mu$}
 \Oval(60,20)(16,20)(0)
 \ArrowLine(60.01,36)(59.01,36)
 \Text(63,36)[bl]{$\bar\nu_e$}
 \ArrowLine(59.99,4)(60.01,4)
 \Text(60.8,7)[bl]{$e^-$}
 \PhotonArc(80,29.2)(29.2,238,341){1.6}{11.5}
 \PhotonArc(81.6,23.2)(14,246,343){1.6}{5.5}
 \Text(70,-12)[t]{(a)}
\end{picture}
\hfill
\begin{picture}(120,70)(0,0)
 \SetWidth{1.5}
 \Line(15,20)(30,20)
 \Line(70,20)(120,20)
 \SetWidth{0.8}
 \ArrowLine(15,20)(30,20)
 \ArrowLine(70,20)(86,20)
 \ArrowLine(86,20)(104,20)
 \ArrowLine(104,20)(120,20)
 \SetWidth{0.5}
 \ArrowLine(30,20)(70,20)
 \Oval(50,20)(16,20)(0)
 \ArrowLine(50.01,36)(49.01,36)
 \ArrowLine(49.99,4)(50.01,4)
 \PhotonArc(61.1,24)(24,238,350){1.6}{10.5}
 \PhotonArc(78,31.2)(26.8,238,335){1.6}{10.5}
 \Text(70,-12)[t]{(b)}
\end{picture}
\hfill\null\\
\hfill
\begin{picture}(120,70)(0,0)
 \SetWidth{1.5}
 \Line(15,20)(30,20)
 \Line(70,20)(120,20)
 \SetWidth{0.8}
 \ArrowLine(15,20)(30,20)
 \ArrowLine(70,20)(84,20)
 \ArrowLine(84,20)(97,20)
 \ArrowLine(97,20)(107,20)
 \ArrowLine(107,20)(120,20)
 \SetWidth{0.5}
 \ArrowLine(30,20)(70,20)
 \Oval(50,20)(16,20)(0)
 \ArrowLine(50.01,36)(49.01,36)
 \ArrowLine(49.99,4)(50.01,4)
 \PhotonArc(94.8,18)(12,15,166){1.6}{6.5}
 \PhotonArc(66,36)(34,250,331){1.6}{10.5}
 \Text(70,-12)[t]{(c)}
\end{picture}
\hfill
\begin{picture}(120,70)(0,0)
 \SetWidth{1.5}
 \Line(20,20)(52.8,20)
 \Line(86.8,20)(122,20)
 \SetWidth{0.8}
 \ArrowLine(20,20)(35,20)
 \ArrowLine(35,20)(52.8,20)
 \ArrowLine(86.8,20)(106,20)
 \ArrowLine(106,20)(122,20)
 \SetWidth{0.5}
 \ArrowLine(53.2,20)(86.8,20)
 \Oval(70,20)(12,16.8)(0)
 \ArrowLine(70.01,32)(69.01,32)
 \ArrowLine(69.99,8)(70.01,8)
 \PhotonArc(60.8,24)(27.2,190,270){1.6}{7.0}
 \PhotonArc(60.8,22)(25.2,270,340){1.6}{5.5}
 \PhotonArc(79.2,22)(25.2,200,270){1.6}{6.0}
 \PhotonArc(79.2,24)(27.2,270,351){1.6}{7.5}
 \Text(70,-12)[t]{(d)}
\end{picture}
\hfill\null\\
\vglue 18pt
\end{center}
\caption{Examples of diagrams whose cuts give contributions to
        $\mu^-\rightarrow\nu_\mu e^-\bar\nu_e$,
        $\mu^-\rightarrow\nu_\mu e^-\bar\nu_e\gamma$ or
        $\mu^-\rightarrow\nu_\mu e^-\bar\nu_e\gamma\gamma$.}

\label{fig:PhotonLoopDiags}
\end{figure}

The result obtained for the complete set of photonic diagrams is
\begin{eqnarray}
\Delta\Gamma_{\gamma\gamma}^{(2)}&=&
\Gamma_0\left(\frac{\alpha_r)}{\pi}\right)^2
\bigg(\frac{11047}{2592}-\frac{1030}{27}\zeta(2)
                         -\frac{223}{36}\zeta(3)
                         +\frac{67}{8}\zeta(4)
                         +53\zeta(2)\ln (2)\bigg)\nonumber\\ \\
                         &=&\Gamma_0\left(\frac{\alpha_r}{\pi}\right)^2
                            3.55877
\end{eqnarray}
where $\zeta(3)=1.20206...$ and $\zeta(4)=\pi^4/90$.

\subsection{Electron-Loops and $e^+e^-$ Pair Creation}

Diagrams containing an electron loop whose cuts give contributions to
muon decay are shown in Fig.3. The result obtained for these
diagrams is

\begin{figure}
\begin{center}
\hfill
\begin{picture}(120,70)(0,0)
 \SetWidth{1.5}
 \Line(20,20)(36,20)
 \Text(20,23)[bl]{$\mu^-$}
 \Line(36,20)(52.8,20)
 \Line(87.2,20)(120,20)
 \SetWidth{0.8}
 \ArrowLine(20,20)(36,20)
 \ArrowLine(36,20)(52.8,20)
 \ArrowLine(87.2,20)(104,20)
 \ArrowLine(104,20)(120,20)
 \SetWidth{0.5}
 \ArrowArc(70,-3.2)(6.4,0,180)
 \Text(68,0)[br]{$e^+$}
 \ArrowArc(70,-3.2)(6.4,180,0)
 \Text(75,-6)[tl]{$e^-$}
 \ArrowLine(50,20)(90,20)
 \Text(70.8,20.2)[bl]{$\nu_\mu$}
 \Oval(70,20)(12,16.8)(0)
 \ArrowLine(70.01,32)(69.01,32)
 \Text(73,33)[bl]{$\bar\nu_e$}
 \ArrowLine(69.99,8)(70.01,8)
 \Text(70.8,11)[bl]{$e^-$}
 \PhotonArc(64,26)(30,191,270){1.6}{9.5}
 \PhotonArc(76,26)(30,270,349){1.6}{9.5}
 \Text(70,-17)[t]{(a)}
\end{picture}
\hfill
\hglue -0.2cm
\begin{picture}(120,70)(0,0)
 \SetWidth{1.5}
 \Line(15,20)(30,20)
 \Line(70,20)(120,20)
 \SetWidth{0.8}
 \ArrowLine(15,20)(30,20)
 \ArrowLine(70,20)(108,20)
 \ArrowLine(108,20)(120,20)
 \SetWidth{0.5}
 \ArrowArc(83.2,3.2)(6.4,0,180)
 \ArrowArc(83.2,3.2)(6.4,180,0)
 \ArrowLine(30,20)(70,20)
 \Oval(50,20)(16,20)(0)
 \ArrowLine(49.99,4)(50.01,4)
 \ArrowLine(50.01,36)(49.99,36)
 \PhotonArc(76,36)(34,242,272){1.6}{4.0}
 \PhotonArc(76,36)(34,292,330){1.6}{4.5}
 \Text(70,-17)[t]{(b)}
\end{picture}
\hfill\null\\
\hfill
\begin{picture}(120,70)(0,0)
 \SetWidth{1.5}
 \Line(20,20)(52.8,20)
 \Line(87.2,20)(120,20)
 \SetWidth{0.8}
 \ArrowLine(20,20)(52.8,20)
 \ArrowLine(87.2,20)(120,20)
 \SetWidth{0.5}
 \ArrowArc(70,-2.8)(6.4,0,180)
 \ArrowArc(70,-2.8)(6.4,180,0)
 \ArrowLine(50,20)(90,20)
 \Oval(70,20)(12,16.8)(0)
 \ArrowLine(69.99,8)(70.01,8)
 \ArrowLine(70.01,32)(69.99,32)
 \PhotonArc(68,8)(12,160,245){1.6}{4.5}
 \PhotonArc(72,8)(12,295,20){1.6}{4.5}
 \Text(70,-17)[t]{(c)}
\end{picture}
\hfill
\begin{picture}(120,70)(0,0)
 \SetWidth{1.5}
 \Line(15,20)(30,20)
 \Line(64.4,20)(125,20)
 \SetWidth{0.8}
 \ArrowLine(15,20)(30,20)
 \ArrowLine(64.4,20)(75.1,20)
 \ArrowLine(75.1,20)(109.3,20)
 \ArrowLine(109.3,20)(125,20)
 \SetWidth{0.5}
 \ArrowArc(92.2,9.3)(6.4,0,180)
 \ArrowArc(92.2,9.3)(6.4,180,0)
 \ArrowLine(30,20)(64.4,20)
 \Oval(47.2,20)(12,16.8)(0)
 \ArrowLine(47.19,8)(47.21,8)
 \ArrowLine(47.21,32)(47.19,32)
 \PhotonArc(85.8,20)(10.7,180,270){1.6}{4.5}
 \PhotonArc(98.6,20)(10.7,270,0){1.6}{4.5}
 \Text(70,-17)[t]{(d)}
\end{picture}
\hfill\null\\
\vglue 18pt
\end{center}
\caption{Diagrams containing an electron loop whose cuts give
         contributions to muon decay,
         $\mu^-\rightarrow\nu_\mu e^-\bar\nu_e$,
         $\mu^-\rightarrow\nu_\mu e^-\bar\nu_e\gamma$ or
         $\mu^-\rightarrow\nu_\mu e^-\bar\nu_e e^+e^-$.}
\label{fig:ELoopDiags}
\end{figure}
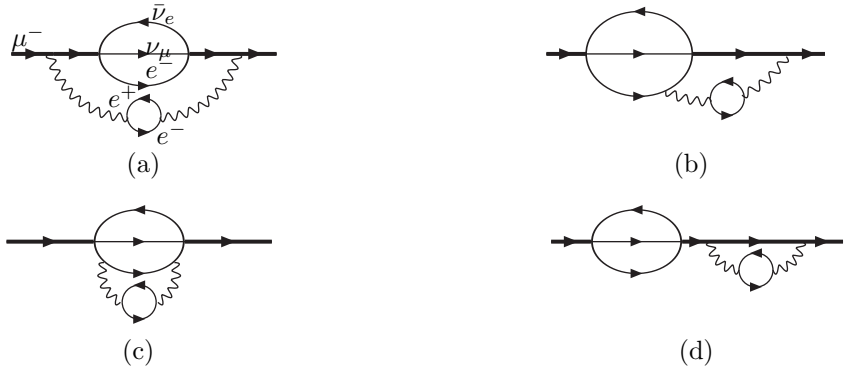

\begin{eqnarray}
\Delta\Gamma_{{\rm elec}}^{(2)}&=&
-\Gamma_0\left(\frac{\alpha_r}{\pi}\right)^2
\left(\frac{1009}{288}-\frac{77}{36}\zeta(2)
                      -\frac{8}{3}\zeta(3)\right)\\
                &=&\Gamma_0\left(\frac{\alpha_r}{\pi}\right)^2
                            3.22034.
\label{eq:electronnum}
\end{eqnarray}

The value given in Eq.(\ref{eq:electronnum})
is consistent with a numerical study carried out by
Luke {\it et al.}\cite{LukeSavaWise} in the context of semi-leptonic
decays of heavy quarks.

In order to obtain a UV finite answer the a diagrams in which
the electron loop is replaced by
the photon 2-point counterterm must be included and therefore a
decision has to taken as to the renormalization scheme that is to be
adopted. This will be discussed further in section~IV.

\section{The Renormalized Electromagnetic Coupling Constant, $\alpha_r$}

The use of dispersion relations to calculate the hadronic and muon loop
contributions in the previous section naturally invokes a subtraction of
the photon vacuum polarization at $q^2=0$ and is therefore equivalent to
the on-shell renormalization scheme. In cases where there are two or
more widely separated scales, such as $m_e$ and $m_\mu$, use of the
$\overline{{\rm MS}}$ renormalization scheme is indicated since it
automatically incorporates the large logarithms that arise
into the value of the renormalized coupling constant, $\alpha_r$,
at tree level.

It is therefore
appropriate here to adopt the $\overline{{\rm MS}}$ renormalization
scheme. The hadronic contributions of section III.A that were obtained
via dispersion relations must be corrected to convert them from the
on-shell
to $\overline{{\rm MS}}$ renormalization scheme. As it turns out the
contribution from muon loops is the same in both schemes when the
't~Hooft mass is taken set to $\mu=m_\mu$ as is appropriate here.
It can be shown\cite{howto} that the
$\overline{{\rm MS}}$ renormalization scheme is implemented in a
consistent manner by using the results of section III.A as they are
given and setting
\begin{equation}
\alpha_r=\alpha_e(m_\mu)\equiv\frac{\alpha}
    {1-\frac{\displaystyle \alpha}{\displaystyle 3\pi}
       \ln\frac{\displaystyle m_\mu^2}{\displaystyle m_e^2}}
                 +\frac{\alpha^3}{4\pi^2}\ln\frac{m_\mu^2}{m_e^2}.
\label{eq:alphaefinal}
\end{equation}
where the logarithm of ${\cal O}(\alpha^3)$ was first calculated
by Jost and Luttinger\cite{JostLuttinger}.
The substitution (\ref{eq:alphaefinal}) correctly resums logarithms of
the form
$\alpha^n\ln^{n-1}(m_\mu^2/m_e^2)$ for all $n>0$ and incorporates those
of $\alpha^3\ln(m_\mu^2/m_e^2)$. Upon evaluation
\[
\alpha_e(m_\mu)=1/135.90=0.0073582.
\]

\section{Conclusions}

It has been over 40 years since the 1-loop QED corrections to the muon
lifetime were calculated. The 2-loop contributions have had to await
the development of new theoretical techniques, as well substantial
increases in computer speed and storage capacity, but are now available.

The complete 2-loop QED contribution to the muon lifetime in the Fermi
model may be encapsulated in the quantity $\Delta q^{(2)}$, as defined
in Eq.s(\ref{eq:QEDcorr}) and (\ref{eq:DeltaqSeries}). After including
the effects of virtual and real photons, virtual electrons, muons, taus
and hadrons as well as $e^+e^-$ pair creation is found to be
\begin{eqnarray}
\Delta q^{(2)}&=&\left(\frac{\alpha_e(m_\mu)}{\pi}\right)^2
        \bigg(\frac{156815}{5184}
             -\frac{1036}{27}\zeta(2)
                         -\frac{895}{36}\zeta(3)
                         +\frac{67}{8}\zeta(4)
                         +53\zeta(2)\ln2
                         -(0.042\pm0.002)
                          \bigg)\\
              &=&\left(\frac{\alpha_e(m_\mu)}{\pi}\right)^2
                      (6.700\pm0.002)
\label{eq:numericalvalue}
\end{eqnarray}
The tiny effect of tau loops has been included in the numerical value
given in eq.(\ref{eq:numericalvalue}).

This translates into a new value for the Fermi coupling constant of
\begin{xxalignat}{2}
{}\qquad\qquad\qquad\qquad
G_F&=(1.16637\pm0.00001)\times10^{-5}\,{\rm GeV^{-2}}&   (9\,{\rm ppm})&
\end{xxalignat}
The error has been halved relative to its previous value and is
now entirely experimental.

New measurements of the muon lifetime are planned at the
Brookhaven National Laboratory, the Paul Scherrer Institute and
the Rutherford-Appleton Laboratory and
it is therefore likely that the uncertainty on $G_F$ from this
source will be reduced to somewhere in the range 0.5--1\,ppm.

In that case the theoretical error should still
be negligible but other issues
such, as error on the muon mass, $m_\mu$, and the upper limit on the
muon neutrino mass, $m_{\nu_\mu}$, need to be considered.

Finally many of the results and techniques employed here can be readily
taken over and applied to inclusive decays of the $b$-quark.

\end{document}